\documentclass[twocolumn,superscriptaddress,showpacs,preprintnumbers,amsmath,amssymb,oneside,prl]{revtex4}

\usepackage{graphicx}
\usepackage{dcolumn}
\usepackage{bm}

\begin{document}

\title{Entangling Color-Different Photons via Time-Resolved Measurement and Active Feed-Forward}

\author{Tian-Ming Zhao}
\affiliation{Hefei National Laboratory for Physical Sciences at Microscale and Department
of Modern Physics, University of Science and Technology of China, Hefei,
Anhui 230026, China}
\affiliation{Synergetic Innovation Center of Quantum Information and Quantum Physics, University of Science and Technology of China, Hefei, Anhui 230026, China}

\author{Han Zhang}
\affiliation{Hefei National Laboratory for Physical Sciences at Microscale and Department
of Modern Physics, University of Science and Technology of China, Hefei,
Anhui 230026, China}
\affiliation{Synergetic Innovation Center of Quantum Information and Quantum Physics, University of Science and Technology of China, Hefei, Anhui 230026, China}

\author{Jian Yang}
\affiliation{Hefei National Laboratory for Physical Sciences at Microscale and Department
of Modern Physics, University of Science and Technology of China, Hefei,
Anhui 230026, China}
\affiliation{Synergetic Innovation Center of Quantum Information and Quantum Physics, University of Science and Technology of China, Hefei, Anhui 230026, China}

\author{Zi-Ru Sang}
\affiliation{Hefei National Laboratory for Physical Sciences at Microscale and Department
of Modern Physics, University of Science and Technology of China, Hefei,
Anhui 230026, China}
\affiliation{Synergetic Innovation Center of Quantum Information and Quantum Physics, University of Science and Technology of China, Hefei, Anhui 230026, China}

\author{Xiao Jiang}
\affiliation{Hefei National Laboratory for Physical Sciences at Microscale and Department
of Modern Physics, University of Science and Technology of China, Hefei,
Anhui 230026, China}
\affiliation{Synergetic Innovation Center of Quantum Information and Quantum Physics, University of Science and Technology of China, Hefei, Anhui 230026, China}

\author{Xiao-Hui Bao}
\affiliation{Hefei National Laboratory for Physical Sciences at Microscale and Department
of Modern Physics, University of Science and Technology of China, Hefei,
Anhui 230026, China}
\affiliation{Synergetic Innovation Center of Quantum Information and Quantum Physics, University of Science and Technology of China, Hefei, Anhui 230026, China}

\author{Jian-Wei Pan}
\affiliation{Hefei National Laboratory for Physical Sciences at Microscale and Department
of Modern Physics, University of Science and Technology of China, Hefei,
Anhui 230026, China}
\affiliation{Synergetic Innovation Center of Quantum Information and Quantum Physics, University of Science and Technology of China, Hefei, Anhui 230026, China}

\begin{abstract}
Entangling independent photons is not only of fundamental interest but also of crucial importance for quantum information science. Two-photon interference is a major method to entangle independent identical photons. If two photons are color-different, perfect two-photon coalescence cannot happen anymore, which makes the entangling of color-different photons difficult to realize. In this letter by exploring and developing time-resolved measurement and active feed-forward, we have entangled two independent photons of different colors for the first time. We find that entanglement with a varying form can be identified for different two-photon temporal modes through time-resolved measurement. By using active feed-forward we are able to convert the varying entanglement into uniform. Adopting these measures, we have successfully entangled two photons with a frequency separation of 16 times larger than their linewidths. In addition to its fundamental interest, our work also provides an approach to solve the frequency mismatch problem for future quantum networks.
\end{abstract}


\maketitle


Entangling independent photons through two-photon interference \cite{Ou1988a, Shih1988a} is ubiquitous in photonic quantum information experiments \cite{Kok2007a,Sangouard2011,Pan2012}. When two identical photons are superimposed on a beam-splitter, the probabilities of both photons are transmitted or both are reflected interfere with each other and result in two-photon coalescence. Such a two-photon interference effect has been first observed by Hong, Ou and Mandel \cite{Hong1987}. From a more fundamental point of view, this interference is due to the bosonic nature of photons \cite{Fedrizzi2008}. Two identical photons have a symmetric wave function, thus their spacial wave function has to be symmetric, which leads to photon coalescence after passing through a beam-splitter. Therefore, only an anti-symmetric two-photon state will lead to a coincidence between different output ports of a beam-splitter, which constitutes the physical basis of measuring Bell states and entangling independent photons. What if the input photons are color-different, can we still make Bell-state measurement and entangle independent photons as usual, for instance in the degrees of polarization, time-bin and momentum?

Entangling color-different photons also has strong practical applications. In quantum networking \cite{Kimble2008}, photons from separate quantum systems are often different in color due to various reasons. For instance, in the condensed matter systems such as quantum dots, nitrogen vacancy (NV) centers, photons from two separate emitters are usually different in frequency due to their different local environments \cite{Patel2010,Bernien2012a}. For all quantum systems, when they are moving with a high-speed (e.g. in a satellite or an airplane), the Doppler effect will give rise to significant frequency shifts for the emitting photons. Besides, the strong interest in hybrid quantum networking by combining the advantages of each physical system also necessitates the entangling operation between color-different photons. Preliminary studies have been carried out on the quantum beat of two color-different photons both theoretically \cite{Legero2003b,Metz2008} and experimentally \cite{Legero2004,Fedrizzi2008}. Without making use of time-resolved measurement \cite{Fedrizzi2008}, rather poor interference visibility has been observed. By making use of time-resolved measurement \cite{Legero2004}, high-visibility interference shows up in a time-dependent fast-oscillating manner, however perfect two-photon coalescence could only happen through narrow temporal filtering. In addition, interfering color-different photons through a dual Mach-Zehnder interferometer has also been studied previously and interesting interference patterns were reported \cite{Larchuk1993}.

In this letter, we study the entangling of independent color-different photons. In order to erase the frequency-distinguishable information during two-photon interference, we make use of time-resolved measurements. Entanglement states can be selected out from an otherwise mixed state for each combination of temporal modes. In our experimental demonstration, we start from two pairs entangled photons which are color-different and make use the entanglement swapping process. In contrast to prior experiments, a time-resolved Bell-state analyzer is utilized and reveals a random phaseshift for the two photons after entanglement swapping. The random phaseshift is later compensated through active feed-forward with a Pockels cell. By taking these measures, we have successfully entangled two independent photons with a frequency separation of 80 MHz which is 16 times larger than their frequency linewidths. With photon detectors of much better time resolution, independent photons with much larger frequency separation can become entangled with our method.

\begin{figure*}[hbtp]
\centering
\includegraphics[width=0.65\textwidth]{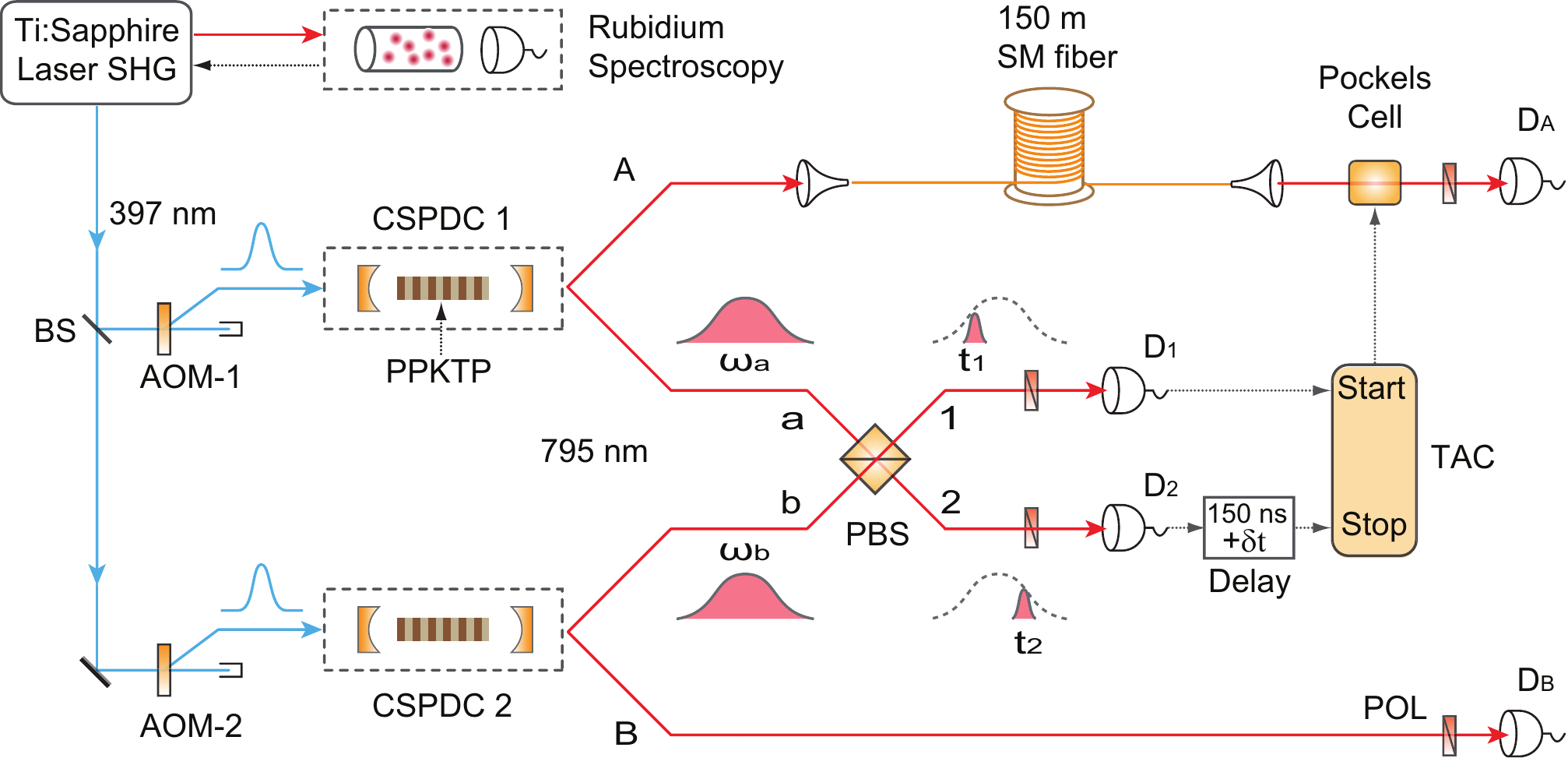}
\caption{(color online). Experimental setup. Two cavity-enhanced spontaneous parametric down-conversion (CSPDC) sources are being used to create two pairs of entangled photons. The pumping beam is generated though second harmonic generation (SHG) of a Ti:Sapphire laser working at 795 nm and stabilized through Rubidium spectroscopy. Two acousto-optic modulators (AOM) are used to chop the pumping beam into short pulses with a repetition rate of 2 MHz and to tune the frequencies of the narrowband entangled photons. Each entanglement source is basically made up of a linear cavity with a 25 mm long nonlinear crystal (PPKTP) and a 5 mm long KTP crystal inside. The KTP is utilized to achieve double-resonance for both CSPDC photons through temperature-tuning \cite{Yangjian2012}. The measured linewidths for the two cavities are $\gamma_{1}/2\pi=4.2$ MHz and $\gamma_{2}/2\pi=5.6$ MHz respectively. Within each source, by controlling the double-resonance condition and making use of additional filtering etalons \cite{Zhang2011}, the two down-converted photons are configured to have the same frequency which is exactly  one half of the pumping beam. Photon $A$ and $a$ are polarization entangled, and so are $B$ and $b$. A polarizing beam-splitter (PBS) along with two polarizers (POL) are utilized for the Bell state measurement. Detection-time differences between $D_{1}$ and $D_{2}$ are fed-forward to a Pockels cell to cancel the random phaseshifts in order to recover the entanglement between photon $A$ and $B$. To compensate the feedback delay due to the single-photon detectors, the time-to-amplitude convertor (TAC) and the high-voltage driver of the Pockels cell, a single-mode (SM) fiber loop with a length of 150 m is inserted for photon $A$.}
\label{setup}
\end{figure*}

We consider two single-photons both of which are initialized in the state of $1/\sqrt{2}(|H\rangle+|V\rangle)$ where $H$ refers to horizontal polarization and $V$ refers to vertical polarization. Ideally when these two photons are indistinguishable with each other, after passing through a polarizing beam-splitter (PBS) which transmits horizontal polarization and reflects vertical polarization, they will become entangled in a state of $1/\sqrt{2}(|H\rangle_1|H\rangle_2-|V\rangle_1|V\rangle_2)$ with the subscripts $1$ and $2$ denoting the two output ports of the PBS shown in Fig. \ref{setup}, if we only consider the case that two photons exit from different ports. While if the input two photons are color-different, say, photon $a$ has a frequency of $\omega_a$ and photon $b$ has a frequency of $\omega_b$, the output two-photon state will change to $|\psi\rangle_{12}=1/\sqrt{2}(|H,\omega_b\rangle_1|H,\omega_a\rangle_2-|V,\omega_a\rangle_1|V,\omega_b\rangle_2)$, where the polarization degree is coupled with the external degree of frequency. If we consider the polarization degree only, the reduced state is thus a maximally mixed state which has little applications for quantum information. While generally such kind of coupling with external degrees or environments can be eliminated using the quantum erasing technique. In order to erase the frequency-distinguishable information, we can make use of fast detections. The output state $|\psi\rangle_{12}$ can be decomposed in the temporal degree with the form of
\begin{align}
|\psi\rangle_{12}= & \frac{1}{\sqrt{2}} \iint d t_1 d t_2[g(t_1)f(t_2)|H,t_1\rangle_1|H,t_2\rangle_2 \\ & -  e^{i(\omega_a-\omega_b)(t_1-t_2)}f(t_1)g(t_2)|V,t_1\rangle_1|V,t_2\rangle_2] \nonumber
\end{align}
Where $f(t)$ and $g(t)$ denotes the temporal shape for the $\omega_a$ photon and the $\omega_b$ photon respectively. If the frequency linewidths of the two photons are similar thus $f(t)\approx g(t)$ and the temporal information is determined much better than $2\pi/|\omega_a-\omega_b|$, the output state will be in an entangled state of $
|\phi\rangle_{12}=1/\sqrt{2}(|H\rangle_1|H\rangle_2-e^{i\Delta\omega\Delta t}|V\rangle_1|V\rangle_2)$ with $\Delta\omega=\omega_a-\omega_b$ and $\Delta t=t_1-t_2$, conditioned on a temporal mode combination of  $|t_1\rangle_1|t_2\rangle_2$. Therefore, a time-resolved measurement enables us to erase the color-different information and select entangled states out of otherwise mixed states.


While determination of the temporal information without affecting the polarization of single-photons requires the high demanding technique of non-demolition measurement \cite{Grangier1998, Nogues1999}, in our experiment we make use the entanglement swapping process \cite{Zukowski1993a, Pan1998, Halder2007} instead. As shown in Fig. \ref{setup}, two pairs of entangled photons are generated from separate sources through cavity-enhanced spontaneous parametric down-conversion (CSPDC) \cite{Ou1999}. The first pair has a frequency of $\omega_a$ and a state of $|\Phi^{+}\rangle_{a}=(1/\sqrt{2})(|H\rangle_{A}|H\rangle_{a}+|V\rangle_{A}|V\rangle_{a})$. The second pair has a frequency of $\omega_b$ and a similar state of $|\Phi^{+}\rangle_{b}=(1/\sqrt{2})(|H\rangle_{B}|H\rangle_{b}+|V\rangle_{B}|V\rangle_{b})$. In order to entangle photon A and photon B, we need to make a joint Bell-state measurement for photon a and b. As shown in Fig. \ref{setup}, Photon a from the first pair and photon b from second pair are superimposed on a PBS, and we only consider the case that two photons leave from different output ports, which gives a final state of
\begin{align}
|\Psi\rangle_f = & 1/\sqrt{2} (|H\rangle_A|H,\omega_b\rangle_1|H,\omega_a\rangle_2|H\rangle_B \nonumber \\
& \,\,\,\,\,\,\,\,\, -|V\rangle_A|V,\omega_a\rangle_1|V,\omega_b\rangle_2|V\rangle_B) \nonumber \\
= & 1/2 \iint d t_1 d t_2 f(t_1)f(t_2) \label{swapping} \\ & \,\,\,\,\,\,[(|H\rangle_{A}|H\rangle_{B}-e^{i\Delta\omega\Delta t}|V\rangle_{A}|V\rangle_{B})\otimes|\Phi^+\rangle_{12} \nonumber \\
& + (|H\rangle_{A}|H\rangle_{B}+e^{i\Delta\omega\Delta t}|V\rangle_{A}|V\rangle_{B})\otimes|\Phi^-\rangle_{12}] \nonumber
\end{align}
where $|\Phi^+\rangle_{12}=1/\sqrt{2}(|H,t_1\rangle_1|H,t_2\rangle_2+|V,t_1\rangle_1|V,t_2\rangle_2)$
can be distinguished unambiguously from a coincidence event of either $|+\rangle_1|+\rangle_2$ or $|-\rangle_1|-\rangle_2$ and $|\Phi^-\rangle_{12}=1/\sqrt{2}(|H,t_1\rangle_1|H,t_2\rangle_2-|V,t_1\rangle_1|V,t_2\rangle_2)$
can be distinguished from a coincidence event of either $|+\rangle_1|-\rangle_2$ or $|-\rangle_1|+\rangle_2$ with $|\pm\rangle=1/\sqrt{2}(|H\rangle\pm|V\rangle)$. Thus a Bell-state measurement result of $|\Phi^\pm\rangle_{12}$
projects the remaining two photons into an entangled state of $|H\rangle_{A}|H\rangle_{B}\mp e^{i\Delta\omega\Delta t}|V\rangle_{A}|V\rangle_{B}$. In contrast to the traditional frequency-degenerate case ($\Delta\omega$=0), conditioned on a Bell-state measurement result, the quantum state of photon $A$ and $B$ after entanglement swapping is not in a definite entangled state anymore, but in an entangled state with its internal phase depending on the detection-time difference and the frequency separation. Since the detection-time difference $\Delta t$ varies from event to event, in average, the final state after entanglement swapping is thus in a mixed state. This problem can be solved by using tight temporal filtering by only selecting the events with $\Delta t \approx 0$, which however will lead to significant reduction of photon flux. A much better method is to make time-resolved measurement and actively compensate the $\Delta t$-dependent phaseshift.

\begin{figure}[hbtp]
\centering
\includegraphics[width=1\columnwidth]{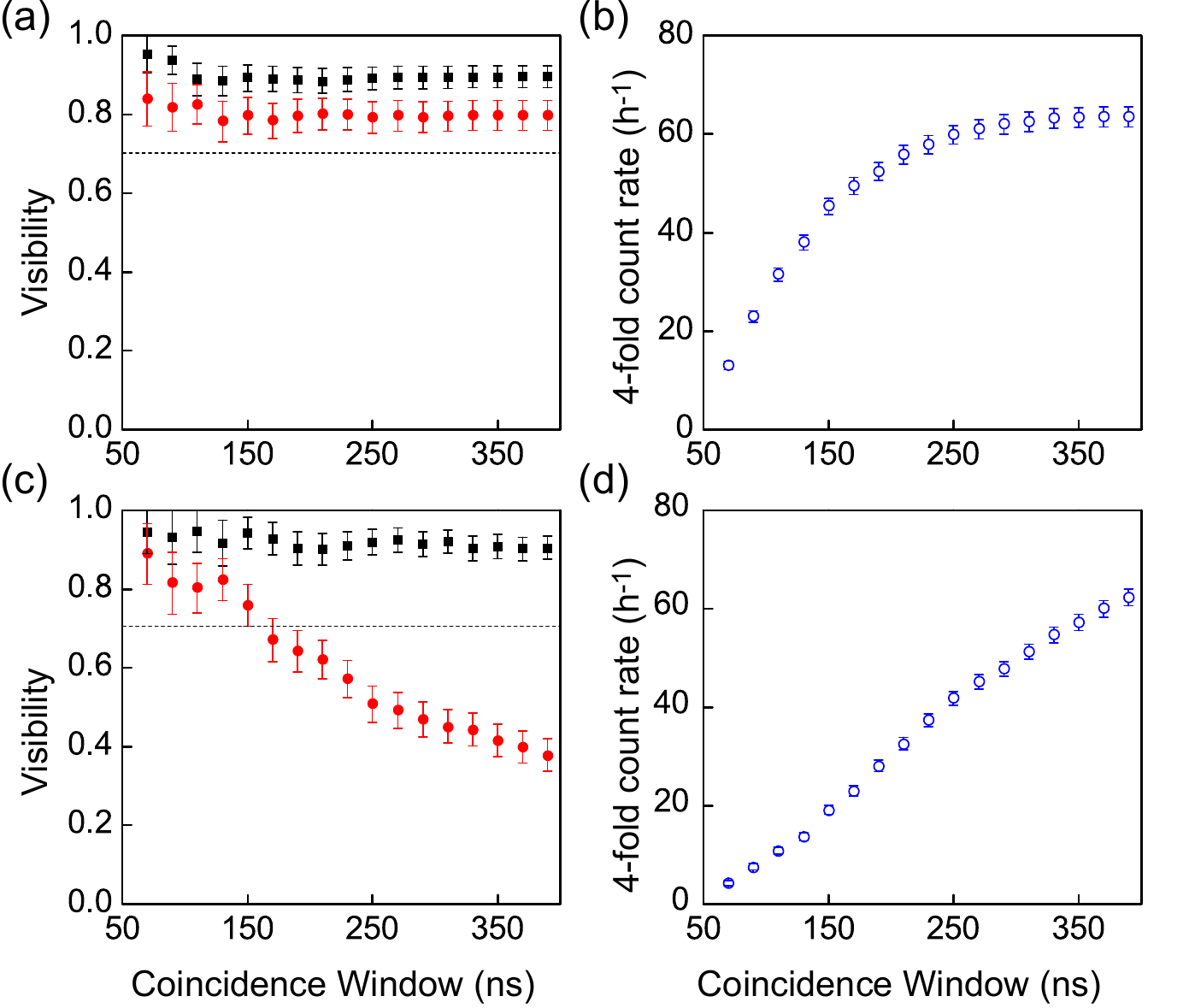}
\caption{(color online). Measured correlation visibilities and 4-fold count rates as a function of the coincidence time window. (a-b), Pumping pulse width is set to 50 ns. (c-d) Pumping pulse width is set to 300 ns. Data points in black filled square are measured in the basis of $|H\rangle/|V\rangle$, and data points in red filled circle are measured in the basis of $|+\rangle/|-\rangle$. In this measurement, 2-fold coincidence count rates are 310 s$^{-1}$ and 190 s$^{-1}$ respectively for the two CSPDC sources. Error bars stand for statistical errors.}
\label{window}
\end{figure}

Prior to study entanglement swapping with color-different photon pairs, we would like to verify that our experimental setup works well for the same frequency case ($\Delta\omega$=0). Photons from CSPDC are usually correlated in frequency \cite{Zhang2011}, which severely limits the fidelity of entanglement swapping with these sources \cite{Zukowski1995}. One solution is to use short pumping pulses with its width smaller than the coherence time for the down-converted photons \cite{Aboussouan2010,Patel2012a}. To verify the elimination of frequency correlation, we measure the polarization-correlation visibilities of photon $A$ and $B$ in the basis of $|H\rangle/|V\rangle$ and $|+\rangle/|-\rangle$ conditioned on two-photon coincidence events of $|+\rangle_{1}|+\rangle_{2}$. Four-fold coincidence counts are analyzed with a commercial multi-channel time analyzer (Agilent U1051A). By setting the pulse width for the pumping beam to 50 ns or 300 ns, we measure the visibilities for different coincidence time windows. For frequency-correlated photon pairs, the polarization visibility after entanglement swapping drops as the coincidence time window goes larger \cite{Zukowski1995}. While for photon pairs without frequency-correlation, the visibility will stay constant. The experimental result is shown in Fig. \ref{window}, which shows clearly that under the pulse width of 50 ns, frequency correlation is eliminated very well. Thus for all remaining measurements, the pulse width for the pump is set to this value. Under the coincidence time window of 300 ns, the two-photon polarization visibilities in the $|H\rangle/|V\rangle$ and $|+\rangle/|-\rangle$ bases are $0.89(3)$ and $0.80(4)$ respectively, which are much higher than the threshold of 0.71 to violate the Bell-CHSH inequality \cite{Clauser1969}. There are several origins for the imperfect visibilities, including slight difference of $f(t)$ and $g(t)$ due to linewidth mismatch between the two CSPDC sources, contribution of multi-pair events from each CSPDC, imperfect spacial overlapping on the PBS, etc.

\begin{figure}[hbtp]
\centering
\includegraphics[width=1\columnwidth]{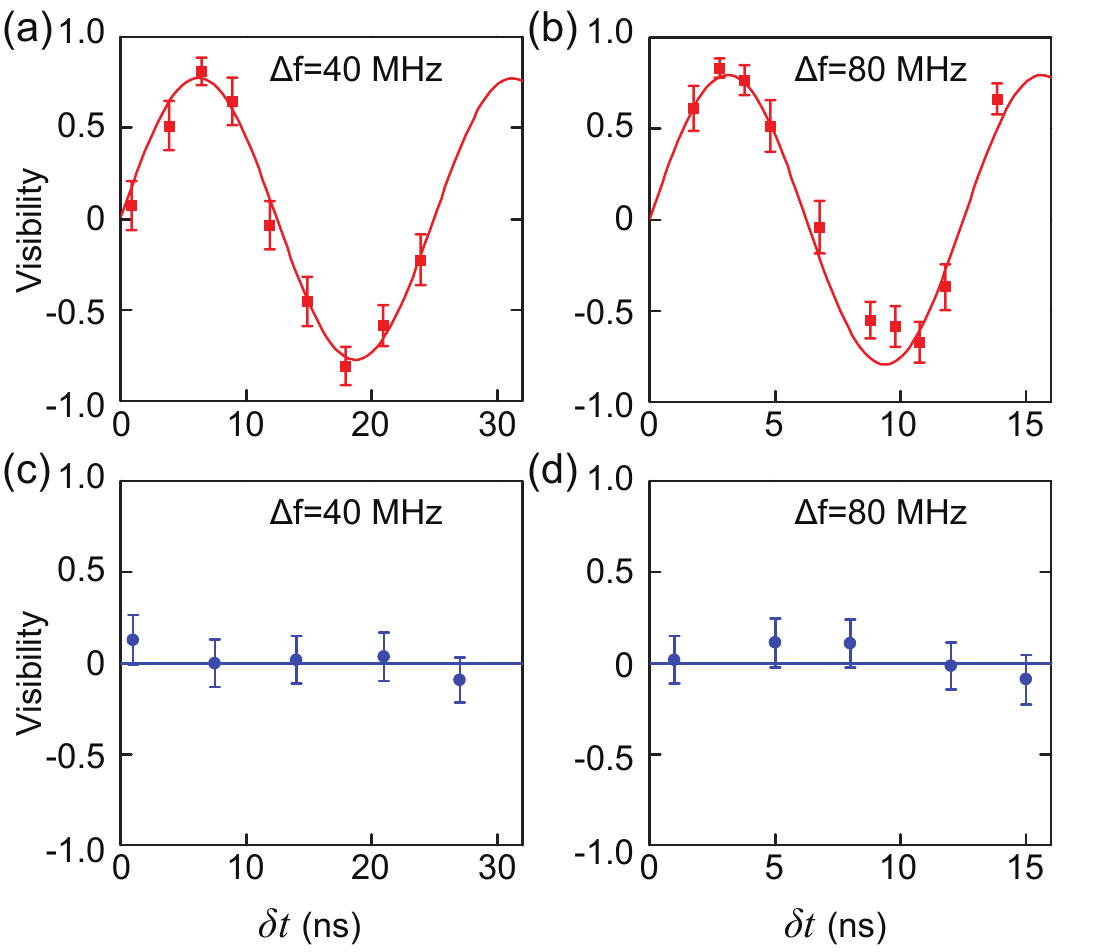}
\caption{(color online). Visibilities in the basis of $|+\rangle/|-\rangle$ for photon $A$ and $B$ as a function of detection time delay $\delta t$. (a-b) Active phase feed-forwarding is applied. (c-d) Active phase feed-forwarding is not applied. Error bars stand for statistical errors. For a given value of $\Delta \omega$, all data points are measured with the same setting for the feed-forward circuitry. In this measurements, 4-fold count rate is 24 h$^{-1}$, and 2-fold count rates are 200 s$^{-1}$ and 140 s$^{-1}$ respectively.}
\label{delay}
\end{figure}

Next we set $\omega_a$ and $\omega_b$ to be different by changing the acousto-optic modulator (AOM) working frequencies for the pumping beams as shown in Fig. \ref{setup}. To compensate the random phaseshift $\Delta\omega\Delta t$ in Eq. \ref{swapping}, we make use of a home-built time-to-amplitude converter (TAC) and a set of fast Pockels cell. The home-built TAC converts the detection-time difference into a voltage amplitude, which is fed into the high-voltage driver for the Pockels cell. The systematic time delay of this feedback system is about 360 ns and we use a fiber loop of 150 m to compensate it. The output state becomes the desired state of $|H\rangle_{A} |H\rangle_{B} \mp |V\rangle_{A} |V\rangle_{B}$ when the phase modulated for photon $A$ by the Pockels cell is equal to $-\Delta\omega\Delta t$. Since the TAC is unipolar and $\Delta t$ fluctuates in the range of $\pm 150 \,\, \rm{ns}$, we add a fixed delay of 150 ns plus an adjustable time delay $\delta t$ for the electronic output signal of $D_{2}$. Thus the actual phase modulated by the Pockels cell is $\Delta\omega(150 \,\, \rm{ns} +\delta t-\Delta t)$. In order to measure the entanglement quality after entanglement swapping and to test the validity of this feed-forward system, we measure the $A$-$B$ correlation visibilities in the basis of $|+\rangle/|-\rangle$ for a series of $\delta t$ points conditioned on the two-photon coincidence events of $|+\rangle_{1}|+\rangle_{2}$. As a comparison, we also make the same measurement for the case without active phase feed-forwarding. Both experimental results are shown in Fig. \ref{delay}. It clearly shows that the $|+\rangle/|-\rangle$ visibilities are recovered by using active phase feed-forward. Fitted visibilities for $\Delta\omega=2\pi\times40$ MHz and $\Delta\omega=2\pi\times80$ MHz are 0.77(4) and 0.80(5) respectively. These results are similar as the $|+\rangle/|-\rangle$ visibilities observed for the same frequency case ($\Delta \omega = 0$), which implies that the main limitations for the visibilities are still due to our imperfect CSPDC sources. By using sources with better lindwidth matching, reducing the excitation probabilities and optimizing the spacial overlapping, these visibilities could increase significantly.

\begin{figure}[hbtp]
\centering
\includegraphics[width=1\columnwidth]{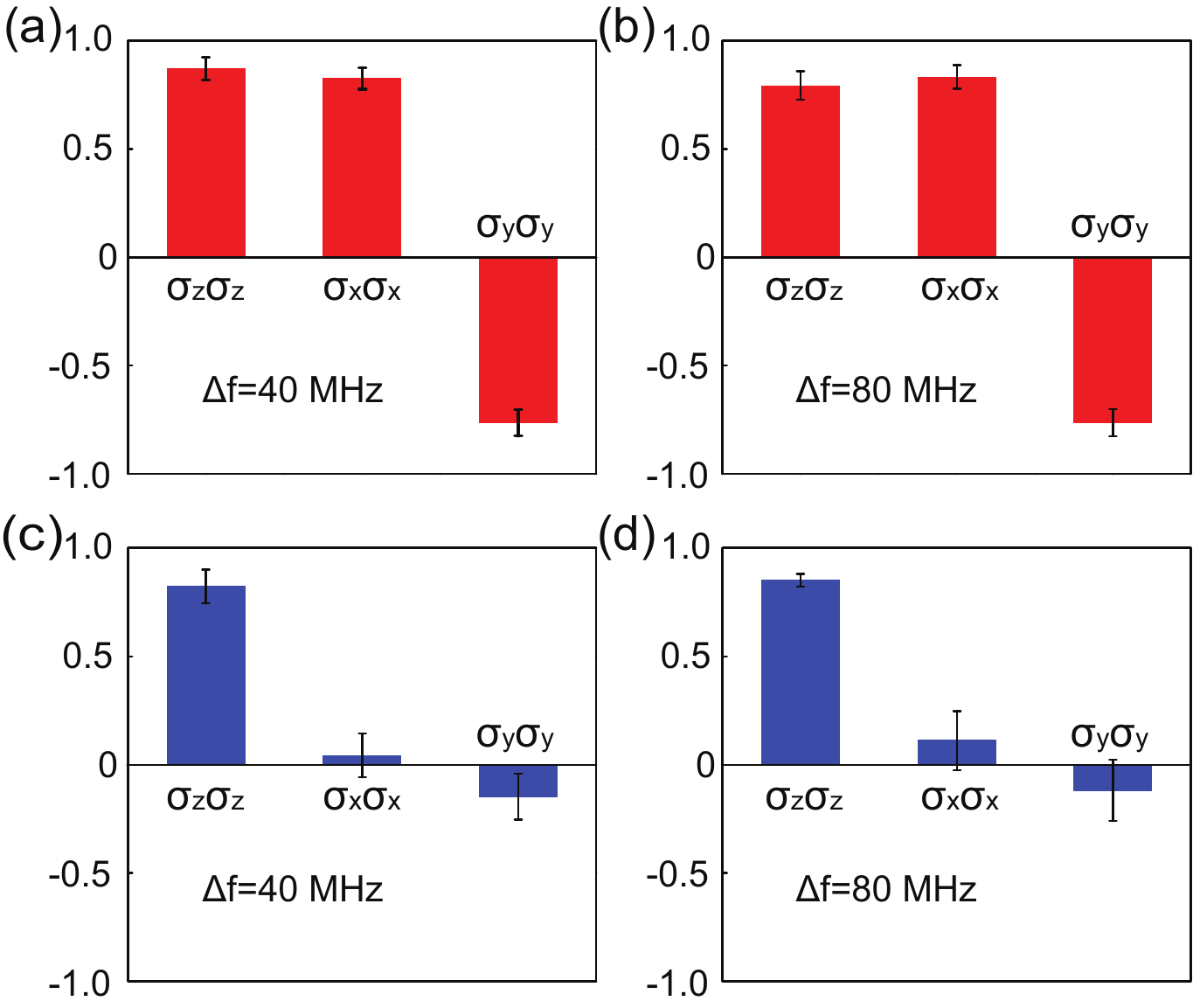}
\caption{(color online). Measurement results of $\sigma_{i}\otimes\sigma_{i}$ for photon $A$ and $B$ with i = x, y and z.  (a-b) Active phase feed-forward is applied. (c-d) Active phase feed-forward is not applied. Error bars stand for statistical errors.}
\label{witness}
\end{figure}

In order to further and directly verify the entanglement after entanglement swapping, we adopt the method of entanglement witness \cite{Guhne2009}. As the desired state of photon $A$ and $B$ is $1/\sqrt{2}(|H\rangle_{A} |H\rangle_{B}+|V\rangle_{A} |V\rangle_{B})$, we select an entanglement witness with the form of $W=1/4(\hat{I}-\sigma_{x}\otimes\sigma_{x}+\sigma_{y}
\otimes\sigma_{y}-\sigma_{z}\otimes\sigma_{z})$ \cite{White2007}. If $W<0$, it implies that photons $A$ and $B$ are in a genuine entangled state \cite{Bourennane2004}. While for a maximally entangled state, $W$ gets its minimum value of $-0.5$. Experimental results on are shown in Fig. \ref{witness}. When applying the active phase feed-forward, the measured
results are $W(\mathrm{40\,MHz})=-0.37(2)$ and $W(\mathrm{80\,MHz})=-0.35(3)$ respectively, both of which clearly prove that photon $A$ and $B$ are genuinely entangled. In comparison, we also measure the $W$ values for the cases without active phase feed-forward, and the results are $W(\mathrm{40\,MHz})=0.00(4)$ and $W(\mathrm{80\,MHz})=0.02(5)$ respectively, which imply that no entanglement can be detected.

While in our experiment since commercial single-photon detectors with moderate time resolution ($\sim$350 ps) have been used, the maximal frequency separation allowed is calculated to be $\sim$630 MHz, which will allow compensating the Doppler effect for an airplane. If state-of-the-art fast electronic single-photon detectors ($\sim$30 ps) \cite{Hadfield2009} are used, entanglement swapping with a frequency separation of $\sim$7.3 GHz can be achieved, which will allow entangling dissimilar NV centers \cite{Bernien2012a} through two-photon interference or compensating the Doppler effect for a satellite. While sum frequency generation with ultrafast laser pulses enables photon detection much faster than electronics, time resolution of $\sim$150 fs has been demonstrated by Kuzucu \textit{et al} \cite{Kuzucu2008}. If such detection technique is utilized, it will enable entanglement swapping with a frequency separation as large as $\sim$1.5 THz, which will allow entangling dissimilar quantum dots \cite{Patel2010} with our method.


In summary, we have experimentally entangled two color-different photons in polarization by time-resolved measurement and active feed-forward. The time-resolved measurement enables us to select entangled states out of otherwise totally mixed states. Active feed-forward enables us to compensate a random phase given from the time-resolved measurement. From a fundamental point of view, our experiment shows that two-photon interference does not only entangle identical photons but also color-different photons. From a practical point of view our experiment provides a ubiquitous approach to solve the frequency-mismatch problem for the interconnection of dissimilar quantum systems, thus may become an essential tool for future quantum networks.

We would like thank to B. Zhao and C.-Y. Lu for helpful discussions. This work was supported by the National Natural Science Foundation of China, National Fundamental Research Program of China (under Grant 2011CB921300), and the Chinese Academy of Sciences. X.-H. B. acknowledges support from and the Youth Qianren Program. H. Z. acknowledges support from China Postdoctoral Science Foundation.

\bibliography{myref}

\end{document}